\documentclass [aps,prb,twocolumn,showpacs]{revtex4}
\usepackage{graphicx}
\usepackage{amsmath}
\usepackage{braket}
\usepackage{xcolor}

\begin{document}
\preprint{Submitted to PRB}
\title{Layered optomagnonic structures: Time-Floquet scattering-matrix approach}
\author{Petros Andreas Pantazopoulos}
\author{Nikolaos Stefanou}
\affiliation{Section of Solid State Physics,
National and Kapodistrian University of Athens, \\
         Panepistimioupolis, GR-157~84 Athens, Greece}
\date{\today}
\begin{abstract}
A fully dynamic theoretical approach to layered optomagnonic structures, based on a time-Floquet scattering-matrix method, is developed. 
Its applicability is demonstrated on a simple design of a dual photonic-magnonic cavity, formed by sandwiching a magnetic garnet thin film between two dielectric Bragg mirrors, subject to continuous excitation of a perpendicular standing spin wave.
Some remarkable phenomena, including nonlinear photon-magnon interaction effects and enhanced inelastic light scattering in the strong-coupling regime, fulfilling a triple-resonance condition, are analyzed and the limitations of the quasi-static adiabatic approximation are established.
\end{abstract}
\maketitle

\section{\label{sc:intro}Introduction}
Propagation of electromagnetic (EM) waves in dynamic media with a
periodic (spatio)temporal variation of their refractive index,
realized, for example, by the action of a progressive sinusoidal
disturbance, have long been investigated and a number of
intriguing effects, such as formation of frequency and wave vector
band gaps, wavelength conversion, pulse shaping, and parametric
amplification, were properly
analyzed.~\cite{morgenthaler,simon,hessel,holberg,harfoush,notomi,biancalana,zurita,koutserimpas}
The lack of time invariance of the linear medium can also lead to
nonreciprocal optical response, thus offering a promising
alternative in the design of compact nonreciprocal photonic
devices (see, e.g., Ref.~\onlinecite{sounas} and references
therein).

Spatiotemporal modulation of the electric permittivity tensor of magnetic materials is induced, for instance, by the various types of spin waves that can be excited in different confined geometries,~\cite{stancil_book} due to the periodic variation of the magneto-optic coupling coefficients,~\cite{hu} causing diverse interesting and useful effects. 
For example, diffraction and mode conversion of guided optical waves, induced by magnetostatic waves in magnetic garnet thin films, offer the potential of large time-bandwidth optical signal processing at microwave frequencies of 1 to 20 GHz and higher.~\cite{fisher,stancil,prabhakar,fetisov}
On the other hand, inelastic light scattering by thermally excited spin waves is widely used for probing magnetic properties of bulk and microstructured materials through Raman or Brillouin spectroscopy.~\cite{fleury,sandercock,grunberg,grimsditch,hillebrands,cochran,jorzick,demokritov,meloche,sebastian}
It should be noted that the magneto-optic interaction is inherently weak and, therefore, the above effects can be in most cases described by first-order perturbation theory.

Greatly increased light--spin-wave interaction phenomena may arise from the simultaneous localization of both fields in the same ultra-small region of space for a long time period. 
This can be achieved, for example, in(sub)millimeter magnetic garnet spheres, which support optical whispering gallery modes~\cite{osada,zhang,haigh1,hisatomi,sharma,haigh2} or Mie resonances,~\cite{almpanis1} as well as in appropriately designed, so-called photomagnonic (or optomagnonic), stratified structures that comprise magnetic dielectric layers and exhibit dual, photonic and magnonic, functionalities~\cite{klos,dadoenkova,pantazopoulos,pantazopoulos2}  combining the properties of photonic~\cite{joannopoulos} and magnonic~\cite{kruglyak2010,zivieri2012} crystals.    
Strong parametric photon-magnon coupling effects, allowing for coherent manipulation of elementary magnetic excitations in solids by optical means and vice versa, have been anticipated~\cite{viola,liu} while large dynamic frequency shift and enhanced modulation of the optical field through multi-magnon absorption and emission processes by a photon have also been reported~\cite{almpanis1,pantazopoulos} in both of the above (spherical and planar) optomagnonic cavity designs.

Since the frequency of a pump spin wave is typically a few orders of magnitude smaller than that of a probe light beam in the infrared and visible range of the EM spectrum, interaction between the two fields can be in principle described by a quasi-static adiabatic approximation. 
In practice, this translates in calculating the relevant optical scattering amplitudes at a sequence of frozen snapshots of the spin wave, during a period, and, at the end of the calculation, obtain the frequency-domain response by Fourier transform.~\cite{almpanis1,pantazopoulos} 
A similar approach was also successfully applied to analyze acousto-optic interaction effects in corresponding phoxonic cavities,~\cite{gantzounis,psarobas,almpanis2,almpanis3} encompassing both weak- and strong-coupling regimes and recovering the results of the widely employed linear photoelastic model in the weak-coupling limit.~\cite{almpanis3}
However, the quasi-static adiabatic approximation has a number of drawbacks, which may be more likely manifested in the case of a spin instead of an acoustic pump wave because the former can attain higher frequencies without being strongly attenuated. 
The adiabatic approximation precludes energy transfer from one wave to the other while, even more importantly, the frequency of the pump wave is decoupled from that of the light beam and thus its actual value becomes immaterial. 
Consequently, this approximation cannot properly describe, e.g., inelastic light scattering processes in a triple-resonance condition, which occur in optomagnonic cavities,~\cite{osada,zhang,haigh1} where the frequency of the magnon matches a photon transition between two resonant modes.

It is therefore tempting to formulate a rigorous, fully dynamic Floquet scattering-matrix method for (spatio)temporal periodic media, adapted to the case of stratified optomagnonic architectures, following, e.g., an approach analogous to that developed in relation to quantum transport in periodically driven systems.~\cite{wagner,li,moskalets,agarwal,bilitewski} 
The purpose of the present paper is to develop such a method and apply it to a specific example of layered optomagnonic structure, establishing the limitations of the quasi-static adiabatic approach. 
The remaining of the paper is structured as follows.
In Sec.~\ref{sc:structure} we introduce our optomagnonic model structure, we present some approximate methods for its theoretical description, namely  the first-order Born approximation  and the adiabatic approach  , and briefly recall the results of the latter in the case under consideration.
In Sec.~\ref{sc:fully_dynamic} we develop a fully dynamic time-Floquet scattering-matrix method and report a detailed study of the given optomagnonic structure, encompassing the quasi-static limit and providing a consistent interpretation of some remarkable effects which cannot be accounted for by the adiabatic approximation.
The last section concludes the paper.

\section{\label{sc:structure}Structure Description and Approximate Treatments} 
\begin{figure}
	\centering
	\includegraphics[width=1\linewidth]{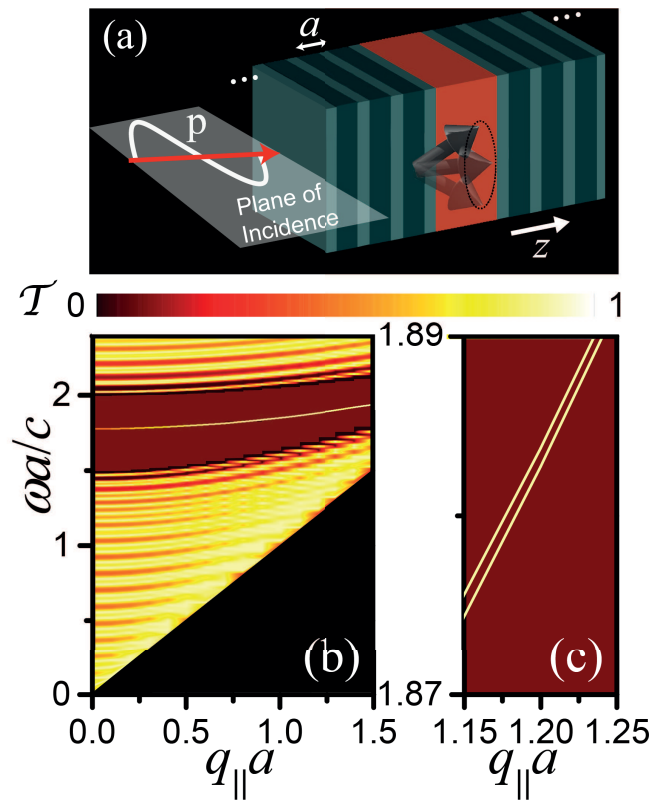}
	\caption{(Color online) (a) Schematic view of a symmetric optomagnonic cavity, formed by a trilayer $\rm TiO_2$/Bi:YIG/$\rm TiO_2$ sandwiched between two $\rm SiO_2/TiO_2$ multilayer Bragg mirrors. 
	Each Bragg mirror consists of 14 periods, with lattice constant $a$, of alternating $\mathrm{SiO}_2$ and $\mathrm{TiO}_2$ layers, of thickness $0.6a$ and $0.4a$, respectively. 
	The total thickness of the trilayer is $1.5a$ and the middle Bi:YIG film, $0.7a$ thick, is magnetically saturated perpendicular to the interfaces. (b) Transmittance of the structure of (a) for p-polarized light incident with given $q_\parallel$. The dark area marks the light cone $\omega=cq_\parallel$, where $\omega$ is the angular frequency and $c$ the speed of light in vacuum. (c)  An enlarged view of (b) in a narrow spectral range within the lowest Bragg gap showing the splitting of the two defect modes.}
	\label{fig1}
\end{figure}
In magnetic materials, the interaction of visible and infrared light with spin waves is inherently weak. 
Optomagnonic cavities aim at enhancing this interaction by simultaneously confining both light and spin waves in the same ultra-small region of space for a long time period. 
In the present paper, following our previous work,~\cite{pantazopoulos} we investigate further a planar optomagnonic cavity formed by a versatile trilayer configuration that comprises a middle magnetic dielectric film made of bismuth-substituted yttrium iron garnet (Bi:YIG) sandwiched between two dielectric Bragg mirrors, as schematically depicted in Fig.~\ref{fig1}(a).
More specifically, we consider a $\mathrm{TiO_2}$/Bi:YIG/$\mathrm{TiO_2}$ trilayer and two symmetric Bragg mirrors, each consisting of 14 periods, with lattice constant $a$, of alternating $\mathrm{SiO}_2$ and $\mathrm{TiO}_2$ layers, of thickness $0.6a$ and $0.4a$, respectively, stacked along the $z$ direction.
The whole structure is embedded in air. 
The trilayer consists of a Bi:YIG film, $0.7a$ thick, magnetically saturated perpendicular to the interfaces, placed between two identical $\mathrm{TiO_2}$ layers, each of thickness $0.4a$.
Assuming $a$ to be of the order of a few hundred nanometers, the operation wavelengths are in the near-infrared region, where the material relative magnetic permeability is equal to unity and Bi:YIG is characterized by a relative electric permittivity tensor of the form
\begin{equation}\label{eq:epsilon}
	\boldsymbol{\epsilon}=\left(
	\begin{array}{ccc} 
	\epsilon& if&0\\
	-if &\epsilon&0\\
	0&0&\epsilon \\
	\end{array} \right)\;
\end{equation}
with $\epsilon=5.5$ and $f=-0.01$.\cite{zvezdin_book} 
$\mathrm{TiO}_2$ and $\mathrm{SiO}_2$ are optically isotropic materials described by a relative permittivity constant equal to 5.35 and 2.13, respectively.~\cite{Schiefke,gao}

The periodic structuring of the Bragg mirrors gives rise to frequency gaps, where light propagation is forbidden.~\cite{joannopoulos}
Figure~\ref{fig1}(b) displays the transmission spectrum of the structure described here above, under illumination by p-polarized light, i.e. light linearly polarized in the plane of incidence, with a wave-vector component parallel to the interfaces $q_\parallel=({q_x^2+q_y^2})^{1/2}$, in a frequency window about the lowest Bragg gap.
We note that, due to translation invariance in the $x-y$ plane, $\mathbf{q}_\parallel$ remains constant.
The presence of the magnetic film affects the optical response of the structure in two ways. 
Firstly, it breaks periodicity leading to the appearance of two resonant modes in the gap, spatially localized in the defect region.~\cite{joannopoulos} 
Secondly, it removes reflection symmetry with respect to the plane of incidence, so that the p and s polarization modes, i.e., modes linearly polarized in and normal to the plane of incidence, respectively, are no longer true eigenmodes of the system.
The defect modes do not have a well-defined linear polarization character and thus both couple, of course to a different degree, to any linearly polarized incident light beam. 
Here, p-polarized light is used to excite these modes, as shown in Fig.~\ref{fig1}(c). 
We note in passing that two distinct modes subsist also in the unmagnetized slab. 
They originate from the splitting of a doubly degenerate mode at normal incidence, where the two polarization degrees of freedom are, obviously, equivalent. 
The difference is that, in the case of the unmagnetized slab, reflection symmetry with respect to the plane of incidence implies a specific linear polarization character, p or s, to each of these two modes at off-normal incidence.

Apart from the localization of the EM field in the magnetic film, the considered optomagnonic structure serves, also, as a cavity for spin waves. 
A general theory of dipole-exchange spin waves in a homogeneous magnetic film of thickness $d$, with out-of-plane arbitrary orientation of the magnetization, has been reported by Kalinikos and Slavin.~\cite{kalinikos1986}
Here, we shall be concerned with perpendicular standing spin waves, assuming the easy axis of magnetization perpendicular to the film, along the $z$ direction. 
If the film is saturated to $M_\mathrm{s}$ by an externally applied static magnetic field $H \widehat{\mathbf{z}}$, under the boundary condition of perfect pinning of the spin waves at the film boundaries we obtain a series of modes at frequencies

\begin{equation}
\label{eq:swfreq}
\Omega_{\kappa}\!=\!\gamma\mu_0 M_\mathrm{s} \!\left[ \frac{H}{M_\mathrm{s}}-1 + \beta +\alpha \left( \frac{\kappa\pi}{d} \right)^2 \right],\kappa=1,2,...,
\end{equation}
where $\gamma$ is the gyromagnetic ratio, $\mu_0$ the magnetic permeability of vacuum, $\alpha$ the exchange constant, and $\beta$ the dimensionless anisotropy coefficient.
 The corresponding magnetization field profiles are given by

\begin{align}
 \mathbf{M}(\mathbf{r},t)/M_\mathrm{s}&=\eta\sin(\kappa\pi z/d)\cos(\Omega_\kappa t)\widehat{\mathbf{x}}\nonumber\\
 &+\eta\sin(\kappa\pi z/d)\sin(\Omega_\kappa t)\widehat{\mathbf{y}}+\widehat{\mathbf{z}} \label{eq:sw_field}~,
\end{align}where  $\eta$ is the relative spin wave amplitude.~\cite{pantazopoulos} Assuming $\eta=0.1$, the magnetization precession angle is about $5^{\mathrm{o}}$, which is a tolerable value for linear spin waves. 

The interaction between light and spin waves enters through the electric permittivity tensor. 
Specifically, the magnetization field given by Eq.~\eqref{eq:sw_field} induces a spatiotemporal perturbation~\cite{pantazopoulos}
\begin{equation}\label{eq:depsilon_decomp}
   \delta{\boldsymbol{\epsilon}}(z,t)=\frac{1}{2}\big[\delta{\boldsymbol{\epsilon}}(z)\exp{(-i\Omega_\kappa t)}+\delta{\boldsymbol{\epsilon}}^\dagger(z)\exp{(i\Omega_\kappa t)}\big]\;
\end{equation}
in the permittivity tensor of the statically magnetized material, where
\begin{equation}\label{eq:depsilon_zfunc}
   \delta{\boldsymbol{\epsilon}}(z)=f\eta\sin(\kappa\pi z/d)
   \left(\begin{array}{ccc}
     0 & 0&1\\
     0 & 0&i\\
     -1&-i&0
   \end{array}\right)\;
\end{equation}
and the dagger denotes hermitian conjugation.

In a simple anisotropic material, the displacement field is related to the electric field through the electric permittivity tensor.
The underlying constitutive relation assuming local response, i.e., that at each point of space the displacement field depends only on the value of the electric field at the same point, reads $\mathbf{D}(\mathbf{r},t)= \epsilon_{0}\int dt' \overline{\boldsymbol{\epsilon}} (t-t')\mathbf{E}(\mathbf{r},t')$, where $\epsilon_0$ is the electric permittivity of vacuum. 
The relative electric permittivity tensor in the time domain, $\overline{\boldsymbol{\epsilon}}$, depends on the difference $t-t'$, as implied by time homogeneity, and is related to the usual electric permittivity tensor $\boldsymbol{\epsilon}$ (in the frequency domain) though a Fourier transform $\boldsymbol{\epsilon}(\omega)=\int dt\overline{\boldsymbol{\epsilon}}(t)\exp(i\omega t)$.
For monochromatic time-harmonic waves of angular frequency $\omega$, of the form $\mathbf{E}(\mathbf{r},t)=\mathrm{Re}[\mathbf{E}(\mathbf{r})\exp(-i\omega t)]$, we obtain $\mathbf{D}(\mathbf{r},t)=\epsilon_{0} \mathrm{Re}[\boldsymbol{\epsilon}(\omega)\mathbf{E}(\mathbf{r})\exp(-i\omega t)]$.
In the presence of a relatively slow (with respect to the period of the EM wave) modulation, $m_{\alpha\beta}(\mathbf{r},t)$, of the elements of the electric permittivity tensor, such as that induced by a spin wave, one can write $\mathbf{E}(\mathbf{r},t) = \mathrm{Re} [\boldsymbol{\mathcal{E}}(\mathbf{r},t) \exp(-i \omega t)]$, where $\boldsymbol{\mathcal{E}}(\mathbf{r},t)$ is a slowly varying envelope function instead of a time-independent phasor. 
Therefore, ignoring the variation of $m_{\alpha \beta}$ and $\boldsymbol{\mathcal{E}}$ in the time integral of the constitutive relation we obtain
\begin{align}
   &D_{\alpha}(\mathbf{r},t) \cong \epsilon_{0} \sum_{\beta}\int dt' \overline{\epsilon}_{\alpha\beta} (t-t')m_{\alpha\beta}(\mathbf{r},t) E_{\beta}(\mathbf{r},t')  
    \nonumber\\
   &\cong \epsilon_{0} \mathrm{Re} \{\sum_{\beta}\epsilon_{\alpha\beta}(\omega) m_{\alpha\beta}(\mathbf{r},t) \mathcal{E}_{\beta} (\mathbf{r},t) \exp(-i \omega t)\} \label{eq:time_varying_Efield} \;,
\end{align}	
i.e., we can write, approximately, the constitutive relation in a mixed time-frequency domain representation in the form of an instantaneous-response, instead of a time-convolution, equation.

Due to the periodic time variation of the electric permittivity tensor of the structure under consideration in the presence of a perpendicular standing spin wave, $\boldsymbol{\mathcal{E}}(\mathbf{r},t)$ can be expanded in a Fourier series as follows: 
$
   \!\boldsymbol{\mathcal{E}}(\mathbf{r},t) =  \sum_{n=0,\pm 1,\ldots} \!\!\!\!\!\mathbf{E}_{n}(\mathbf{r}) \exp(in\Omega_\kappa t) \;,
$
which yields
\begin{equation} \label{eq:fourierdisp} 
   {\mathbf{E}}(\mathbf{r},t) =  \mathrm{Re}\Big\{ \sum_{n=0,\pm 1,\ldots} \mathbf{E}_{n}(\mathbf{r}) \exp[-i(\omega- n\Omega_\kappa )t]\Big\} \;.
\end{equation}
This equation implies that, for an incident wave of angular frequency  $\omega$,  the total outgoing (transmitted plus reflected) field consists, in general, of an infinite number of monochromatic beams with angular frequencies $\omega, \omega \pm
\Omega_\kappa, \omega \pm 2 \Omega_\kappa, \ldots$, which are produced by elastic and inelastic photon scattering that involves absorption and/or emission of zero, one, two, etc. magnons. We note that magnon absorption (emission) processes stem from the first (second) term of Eq.~(\ref{eq:depsilon_decomp}).			

 In the weak-coupling regime, one can restrict to the first-order Born approximation.  In this approximation, the coupling strength associated to the photon-magnon scattering is proportional to the overlap integral $G=\bra{\mathrm{out}}\delta{\boldsymbol{\epsilon}}\ket{\mathrm{in}}$, where $\braket{\alpha\mathbf{r}t|{\mathrm{in}}}=E_\alpha^{\mathrm{in}}(z)\exp[i(\mathbf{q}_\parallel\cdot\mathbf{r}-\omega t)]$ and $\braket{{\mathrm{out}}|\alpha'\mathbf{r}'t'}=E_{\alpha'}^{\mathrm{out}\star}(z)\exp[-i(\mathbf{q}_\parallel'\cdot\mathbf{r}-\omega' t)]$ denote appropriate incoming and outgoing monochromatic time-harmonic waves in the static magnetic layered structure.
Using Eq.~\eqref{eq:depsilon_decomp} we obtain
\begin{equation}
\label{eq:coupling}
G=if\eta\delta(\mathbf{q}_\parallel-\mathbf{q}_\parallel')\big[\delta(\omega-\omega'-\Omega_{\kappa})g_{-}+\delta(\omega-\omega'+\Omega_{\kappa})g_{+}\big]
\end{equation}
where
\begin{align}
   g_{\pm}=\displaystyle\int &dz \sin (\kappa \pi z/d)\Big\{ E_y^\mathrm{out\star}(z)E_z^\mathrm{in}(z)-E_z^\mathrm{out\star}(z)E_y^\mathrm{in}(z) \nonumber \\
   &\pm i\big[E_x^\mathrm{out\star}(z)E_z^\mathrm{in}(z)-E_z^\mathrm{out\star}(z)E_x^\mathrm{in}(z)\big] \Big\}  \label{eq:scatelement}~.
\end{align}

The delta functions in Eq.~\eqref{eq:coupling} express conservation of in-plane momentum and energy in inelastic light scattering processes that involve emission and absorption of one magnon by a photon, as expected in the linear regime.
In our case here, a simple selection rule can be deduced by considering the symmetry of the EM field and the spin wave.
In a structure such as that described in Fig.~\ref{fig1}(a), which has the mirror symmetry with respect to the $x-y$ plane, $\widehat{ \sigma}_z$, the electic field modes, $\mathbf{E}(z)$, are either even or odd functions of $z$.~\cite{pantazopoulos}
On the other hand, since $\widehat{ \sigma}_z\mathbf{E}(z)=E_x(-z)\widehat{ \mathbf{x}}+E_y(-z)\widehat{ \mathbf{y}}-E_z(-z)\widehat{ \mathbf{z}}$, either $E_x,E_y$ are even function of $z$ and $E_z$ is an odd function of $z$ or $E_x,E_y$ are odd function of $z$ and $E_z$ is an even function of $z$. 
Therefore, the integral in Eq.~\eqref{eq:scatelement} vanishes identically, unless the spatial profile of the perpendicular standing spin wave is an odd (even) function of $z$ for transitions between two optical modes of the same (different) symmetry. 
In the present work, since both optical resonances have the same symmetry, we choose a second-order ($\kappa=2$) perpendicular standing spin wave which is an odd function of $z$.

On the other hand, since the spin-wave frequencies are, typically, a few orders of magnitude smaller than those of visible and infrared light, the slow gradual change of the electric permittivity tensor, given by Eq.~\eqref{eq:depsilon_decomp}, can be looked upon as an adiabatic process. 
Therefore, the optical response of the driven system can be described by first evaluating it with the external parameters held fixed and then, at the end of the calculation, allow them to change, in discrete steps within a time period of the spin wave.
This, so-called quasi-static adiabatic, approach, which is not restricted to the weak-coupling regime, has been undertaken in our previous work~\cite{pantazopoulos} and details about its implementation can be found elsewhere.~\cite{pantazopoulos,pantazopoulos2}
Here, we briefly summarize our main results when the higher optical resonance of Fig.~\ref{fig1}(c) for $q_\parallel a=1.2$, at $\omega_{0} a/c=1.88375$, is excited by p-polarized light incident with the given $q_\parallel$, i.e., at an angle of about $40^{\circ}$, on the structure of Fig.~\ref{fig1}(a) in the presence of a second-order ($\kappa=2$) perpendicular standing spin wave with a relative amplitude $\eta=0.1$. 
The continuous variation of the permittivity tensor of the magnetic film, as implied by Eq.~(\ref{eq:depsilon_decomp}), is taken into consideration in the numerical calculation following a discretization approach. Subdividing the Bi:YIG film into 50 homogeneous elementary sublayers, at each of the 60 in total time steps considered, is sufficient to obtain excellent convergence.

When the dynamic magnetization field is switched on, the permittivity tensor and, consequently, the optical transmission spectrum vary periodically in time with the period of the spin wave.
Even though the overall spectral features do not change significantly as time evolves, the oscillation of the optical defect mode with an amplitude $\Delta \omega$ that is of the order of the resonance width is remarkable.~\cite{pantazopoulos} 
The ratio $\Delta\omega/\Gamma$,  where $\Gamma$ is the calculated width of the higher resonance in the optical transmission spectrum for $q_\parallel a=1.2$ [see Fig.~\ref{fig1}(c)] that defines its inverse lifetime, can be adopted as a quantitative measure of the strength of the photon-magnon interaction. 
Here we obtain $\Delta\omega a/c=0.2\times10^{-5}$ and $\Gamma a/c=0.1\times10^{-5}$.
This strong modulation of the optical cavity mode by the perpendicular standing spin wave should be ascribed to the simultaneous concentration of both fields for a long time period in the cavity region. 
In the wave picture, this is manifested as a relatively large-amplitude periodic oscillation of the position of the sharp optical resonance. 
Correspondingly, in the particle picture, strong inelastic light scattering occurs, with considerable probabilities for absorption and emission of many magnons by a photon.
Indeed, the calculated intensities, $\mathcal{I}_n$, of the outgoing (transmitted plus reflected) light beams, reported in Table~\ref{tab:intensities}, are significant also beyond the relatively large first-order ($n= \pm 1$) components.
It is worth-noting that the quasi-static adiabatic approximation precludes energy transfer between the photon and the magnon fields.~\cite{pantazopoulos2}
 Therefore, in the absence of dissipative losses, $\sum_{n=-\infty}^\infty \mathcal{I}_n$ must be equal to unity, because at each frozen snapshot of the spin wave the electric permittivity tensor is Hermitian [see Eqs.~\eqref{eq:epsilon},~\eqref{eq:depsilon_decomp}, and~\eqref{eq:depsilon_zfunc}], which is confirmed by our calculations.
However, even if there is no energy transfer, coupling between the two fields arises through the time variation of the electric permittivity tensor, induced by the spin wave, which affects the  optical response of the structure. 
The small asymmetry between the Stokes ($n>0$) and anti-Stokes ($n<0$) components in Table~\ref{tab:intensities} is due to the lack of time-reversal symmetry of the magnetic structure.~\cite{pantazopoulos}

\begin{table}[!htb]
  \centering
  \begin{tabular}{c| c c c c c c c c c}     \hline
      $n$  & $-4$ & $-3$ &$-2$ &$-1$ &$0$ &$1$ &$2$ &$3$ &$4$  \\ \hline
      $\mathcal{I}_n$ & $0.015$ & $0.025$ &$0.042$ &$0.070$ &$0.648$ &$0.071$ &$0.043$ &$0.025$ &$0.015$ 
    \end{tabular}
    \caption{Intensities of the elastic and the first few inelastic outgoing beams upon excitation of the higher optical resonance at $q_\parallel a=1.2$ in Fig.~\ref{fig1}(c), subject to the action of a second-order perpendicular standing spin wave with a relative amplitude $\eta=0.1$. } \label{tab:intensities}
\end{table}

Though the quasi-static adiabatic approximation is a useful tool for the study of optomagnonic cavities, it has a number of drawbacks (see Appendix).
Namely, as stated above, it precludes energy transfer between the photon and the magnon fields.
Even more importantly, the frequency of the pump spin wave is decoupled from that of the light beam and thus its actual value is immaterial.
Consequently, this approximation cannot properly describe, e.g., inelastic light scattering processes in a triple-resonance condition, which occur in optomagnonic whispering gallery microresonators,~\cite{osada,zhang,haigh1} when the frequency of the magnon matches a photon transition between two resonant modes.

\section{\label{sc:fully_dynamic}Fully Dynamic Description}
\subsection{Formalism} 
Let us first consider a homogeneous medium, characterized by a scalar relative magnetic permeability $\mu$ and a relative time-periodic electric permittivity tensor, $\boldsymbol{\epsilon}(t)=\boldsymbol{\epsilon}(t+T)$. 
If the time variation of the permittivity is very slow compared with the period of an optical wave, assuming slowly varying envelope functions for the electric and magnetic fields, of the form $\mathbf{F}(\mathbf{r},t)=\mathrm{Re}\Big[\boldsymbol{\mathcal{F}}(\mathbf{r},t)\exp(-i\omega t)\Big]$ where $\mathbf{F}=\mathbf{E},\mathbf{H}$, with the help of Eq.~\eqref{eq:time_varying_Efield} Maxwell equations read

\begin{align}
   \nabla \times \boldsymbol{\mathcal{E}}(\mathbf{r},t)\exp(-i \omega t)&=-\mu_0\mu\partial_t\big[\boldsymbol{\mathcal{H}}(\mathbf{r},t)\exp(-i \omega t)\big] \nonumber\\
    \nabla\times\boldsymbol{\mathcal{H}}(\mathbf{r},t)\exp(-i \omega t)&=\epsilon_0\partial_t[\boldsymbol{\epsilon}(t) \boldsymbol{\mathcal{E}}(\mathbf{r},t)\exp(-i \omega t)]\label{eq:maxwell_rt}\;.
\end{align}

The time-periodic variation of the permittivity tensor results in time-periodic envelope functions and, for this reason, Maxwell equations can be solved by expanding these quantities in (truncated) Fourier series. Considering solutions in the form of plane waves with wavevector $\mathbf{q}$, we have

\begin{align}
\boldsymbol{\epsilon}(t)&=\displaystyle\sum_{n=-N}^{N}\boldsymbol{\epsilon}(n)\exp{(in\Omega t)}\nonumber \\
\boldsymbol{\mathcal{E}}(\mathbf{r},t)&=E_0\displaystyle\sum_{n=-N}^{N}{\mathbf{{e}}}(n)\exp{[i(\mathbf{q}\cdot\mathbf{r}+n\Omega t)]}\nonumber \\
\boldsymbol{\mathcal{H}}(\mathbf{r},t)&=\dfrac{E_0}{Z_0}\displaystyle\sum_{n=-N}^{N}{\mathbf{{h}}}(n)\exp{[i(\mathbf{q}\cdot\mathbf{r}+n\Omega t)]}\;, \label{eq:E_planewave}
\end{align}
where $\Omega=2\pi/T$, $Z_0$ is the impedance of free space, $\mathbf{e}$ and $\mathbf{h}$ are polarization vectors chosen such that $\sum_{n=-N}^N \mathbf{e}(n)\cdot\mathbf{e}^{\star}(n)=1$, and  $E_0$ is the field amplitude.
Substituting Eqs.~\eqref{eq:E_planewave} into Eq.~\eqref{eq:maxwell_rt}, we obtain 

\begin{align}
&\mathbf{q} \times {\mathbf{h}}(n)+\dfrac{(\omega-n\Omega)}{c}\!\!\!\displaystyle\sum_{n'=-N}^{N}\!\!\!\boldsymbol{\epsilon}({n-n'}){\mathbf{e}}{(n')}=0  \nonumber \\
&\mathbf{q} \times {\mathbf{e}}(n)-\dfrac{(\omega-n\Omega)}{c}{\mu}\mathbf{h}{(n)}=0\;,
 \label{eq:maxwell_q}
\end{align}
for $n=-N,-N+1,\cdots,N$. 
For given $x$- and $y$-components of the wavevector $\mathbf{q}$,  Eqs.~\eqref{eq:maxwell_q} can be cast in the form of a $6(2N+1)\times6(2N+1)$ linear eigenvalue problem

\begin{align}\label{eq:system1}
\left(
\begin{array}{cc} 
\underline{\underline{\mathbf{k}}}\,\underline{\underline{\boldsymbol{\epsilon}}}&\underline{\underline{\mathbf{C}}}_1\\
-\underline{\underline{\mathbf{C}}}_1&\underline{\underline{\mathbf{k}}}\mu
\end{array} \right)^{-1}
\left(
\begin{array}{cc} 
\underline{\underline{\mathbf{0}}}&\underline{\underline{\mathbf{C}}}_2\\
-\underline{\underline{\mathbf{C}}}_2&\underline{\underline{\mathbf{0}}}
\end{array} \right)
\left(
\begin{array}{c} 
\underline{\mathbf{e}}\\
\underline{\mathbf{h}}\\
\end{array} \!\right)=\dfrac{1}{q_z}\left(
\begin{array}{c} 
\underline{\mathbf{e}}\\
\underline{\mathbf{h}}\\
\end{array} \!\right)\;,
\end{align}
which can be solved by standard numerical algorithms.~\cite{numerical_recipes} 
In Eq.~\eqref{eq:system1}, by a double underscore we denote $3(2N+1)\times 3(2N+1)$ matrices; $\underline{\underline{\mathbf{0}}}$ is the zero matrix, $\underline{\underline{\mathbf{k}}}=\mathrm{diag}(\mathbf{k}_{-N},\mathbf{k}_{-N+1},\cdots,\mathbf{k}_{N})$ with $\mathbf{k}_n=\mathbf{I}(\omega-n\Omega)/c\equiv\mathbf{I}k_n$, and $\underline{\underline{\mathbf{C}}}_{\,i}=\mathrm{diag}(\mathbf{C}_i,\mathbf{C}_i,\cdots,\mathbf{C}_i)$, $i=1,2$, are block diagonal matrices with

\begin{equation}
\mathbf{C}_1=\left( \begin{array}{ccc} 0&0 &q_y\\ 0&0 &-q_x\\ -q_y&q_x &0 \end{array} \right), \mathbf{C}_2=\left( \begin{array}{ccc} 0&1 &0\\ -1&0 &0\\ 0&0 &0 \end{array} \right)\;,
\end{equation}
while $\underline{\underline{\boldsymbol{\epsilon}}}$ is the Toeplitz matrix of the Fourier coefficients of the electric permittivity tensor
 
\begin{equation}
\underline{\underline{\boldsymbol{\epsilon}}}=\left(
\begin{array}{cccc} 
 \boldsymbol{\epsilon}(0) & \boldsymbol{\epsilon}(-1) & \cdots  & \boldsymbol{\epsilon}(-2N)  \\ 
\boldsymbol{\epsilon}(1) & \boldsymbol{\epsilon}(0) & \cdots  & \boldsymbol{\epsilon}(-2N+1)  \\
 \vdots& \vdots &  & \vdots \\ 
\boldsymbol{\epsilon}(2N) & \boldsymbol{\epsilon}(2N-1) & \cdots  & \boldsymbol{\epsilon}(0)  
\end{array} \right)\;.
\end{equation}
Correspondingly $\underline{\mathbf{e}}$, $\underline{\mathbf{h}}$ are $3(2N+1)$-dimensional vectors

\begin{align}
\underline{\mathbf{e}}=
\begin{pmatrix}
 \mathbf{e}(-N)\\ \mathbf{e}(-N+1) \\ \vdots  \\  \mathbf{e}(N)
\end{pmatrix} \;, \quad  \underline{\mathbf{h}}=
\begin{pmatrix}
 \mathbf{h}(-N)\\ \mathbf{h}(-N+1) \\ \vdots  \\  \mathbf{h}(N)
\end{pmatrix} \;.
\end{align}

We characterize the $4(2N+1)$ physically acceptable solutions (non-zero eigenvalues) of Eq.~\eqref{eq:system1} by a superscript $s=+(-)$, which denotes waves propagating or decaying in the positive (negative) $z$ direction, a subscript $p=1,2$ indicating the two eigen-polarizations, and another subscript $\nu=-N,-N+1,\cdots,N$  which labels the different eigenmodes. 
The electric and magnetic field components of these modes, with unit amplitude, wave
vector $\mathbf{q}_{p\nu}^s=\mathbf{q}_\parallel+q_{p\nu z}^s\widehat{\mathbf{z}}$, and polarization eigenvectors ${\mathbf{e}}_{p\nu}^s(n) $ and    ${\mathbf{h}}_{p\nu}^s(n) $ are given by

\begin{align}
\boldsymbol{ \mathcal{E}}_{p\nu}^s(\mathbf{r},t)&=\displaystyle\sum_{n=-N}^{N}\!\! \exp{\big[i(\mathbf{q}^s_{p\nu}\cdot\mathbf{r}+n\Omega t)\big]}{\mathbf{e}}_{p\nu}^s(n)\nonumber \\
\boldsymbol{\mathcal{H}}_{p\nu}^s(\mathbf{r},t)&=\dfrac{1}{Z_0} \!\!\displaystyle\sum_{n=-N}^{N}\!\!\! \exp{\big[i(\mathbf{q}^s_{p\nu}\cdot\mathbf{r}+n\Omega t)\big]}{\mathbf{h}}^s_{p\nu}(n)\label{eq:EHsolution_dynamic} \;.
\end{align}

In the case of static homogeneous media,
Eq.~\eqref{eq:system1} is reduced to a set of $2N+1$ independent eigenvalue equations~\cite{christofi}

\begin{align}\label{eq:system_iso}
\left(
\begin{array}{cc} 
{k}_n\boldsymbol{\epsilon}&\mathbf{C}_1\\
-\mathbf{C}_1&{k}_n\mu\mathbf{I}
\end{array} \right)^{-1}
\left(
\begin{array}{cc} 
\mathbf{0}&\mathbf{C}_2\\
-\mathbf{C}_2&\mathbf{0}
\end{array} \right)
\left(
\begin{array}{c} 
\mathbf{e}\\
\mathbf{h}\\
\end{array} \right)=\dfrac{1}{q_z}\left(
\begin{array}{c} 
\mathbf{{e}}\\
\mathbf{{h}}\\
\end{array} \right),
\end{align}
one for each value of $n=-N,-N+1,\cdots,N$, where $\boldsymbol{\epsilon}$ are the elements of the corresponding static dielectric tensor. 
From the physically acceptable eigenvectors ${\mathbf{e}}_{pn}^s$ and ${\mathbf{h}}_{pn}^s$ of Eqs.~\eqref{eq:system_iso}, we construct the eigenvectors which appear in Eqs.~\eqref{eq:EHsolution_dynamic} as follows: $\mathbf{e}_{p\nu}^s(n)=\delta_{\nu n}{\mathbf{e}}_{pn}^s$, $\mathbf{h}_{p\nu}^s(n)=\delta_{\nu n}\mathbf{h}_{pn}^s$, while the corresponding wave vectors are $\mathbf{q}_{p\nu}^{s}=\mathbf{q}_{pn}^{s}$.

We now consider a planar interface between two different homogeneous,  in general time-periodic with the same periodicity,  media: (1) on the left and (2) on the right side of the interface. 
The interface is perpendicular to the $z$ axis, which is directed from left to right, at $z = 0$. 
Let us assume an eigenmode of the EM field given by Eqs.~\eqref{eq:EHsolution_dynamic}, with amplitude equal to unity, wave vector $\mathbf{q}^{+(1)}_{p'\nu'}$, and polarization described by ${\mathbf{\underline{e}}}^{+(1)}_{p'\nu'}$, ${\mathbf{\underline{h}}}^{+(1)}_{p'\nu'}$, incident on the interface from the left. 
Scattering at the interface gives rise to transmitted and reflected waves with wave vectors $\mathbf{q}^{+(2)}_{p\nu}$ and $\mathbf{q}^{-(1)}_{p\nu}$, $p=1,2$, $\nu=-N,-N+1,\cdots,N$, respectively, characterized by the associated polarization eigenvectors. 
As a result of the common temporal periodicity of both media on either side of the interface  (this comprises the case where the one medium is time-invariant), the Floquet quasi-frequency $\omega$ remains constant, similarly to the Floquet quasi-momentum, or else the Bloch wave vector, when there is spatial periodicity.~\cite{comphy1,comphy2} 
In addition, translation invariance parallel to the interface implies that $\mathbf{q}_\parallel$ remains the same for all of these waves. 
We denote the relative amplitudes of the transmitted and reflected waves (with respect to the amplitude of the incident wave) by $S_{p\nu;p'\nu'}^{++}$ and $S_{p\nu;p'\nu'}^{-+}$, respectively. 
In a similar manner we can proceed for an EM wave incident on the interface from the right and obtain transmitted and reflected waves of relative amplitudes $S_{p\nu;p'\nu'}^{--}$ and $S_{p\nu;p'\nu'}^{+-}$, respectively. 
Continuity of the tangential components of the EM field at the interface, in each case, yields $4(2N+1)$ linear systems, each of dimension $4(2N+1)\times4(2N+1)$, which can be written in compact form as
\begin{widetext}
\begin{align}
&\left(
\begin{array}{cccccccc} 
-{\underline{{e}}}_{1-Nx}^{-(1)}&- {\underline{{e}}}_{2-Nx}^{-(1)}& \cdots& -{\underline{{e}}}_{2Nx}^{-(1)} & {\underline{{e}}}_{1-Nx}^{+(2)}&  {\underline{{e}}}_{2-Nx}^{+(2)}&\cdots &{\underline{{e}}}_{2Nx}^{+(2)} \\
-{\underline{{e}}}_{1-Ny}^{-(1)}&- {\underline{{e}}}_{2-Ny}^{-(1)}& \cdots&- {\underline{{e}}}_{2Ny}^{-(1)} & {\underline{{e}}}_{1-Ny}^{+(2)}&   {\underline{{e}}}_{2-Ny}^{+(2)}&\cdots & {\underline{{e}}}_{2Ny}^{+(2)} \\
-{{\underline{h}}}_{1-Nx}^{-(1)}& -{{\underline{h}}}_{2-Nx}^{-(1)}& \cdots& - {{\underline{h}}}_{2Nx}^{-(1)} &  {{\underline{h}}}_{1-Nx}^{+(2)}&  {{\underline{h}}}_{2-Nx}^{+(2)}&\cdots & {{\underline{h}}}_{2Nx}^{+(2)} \\
-{{\underline{h}}}_{1-Ny}^{-(1)}& -{{\underline{h}}}_{2-Ny}^{-(1)}&\cdots&  -{{\underline{h}}}_{2Ny}^{-(1)} &  {\underline{{h}}}_{1-Ny}^{+(2)}&  {\underline{{h}}}_{2-Ny}^{+(2)}&\cdots &  {{\underline{h}}}_{2Ny}^{+(2)} \\
\end{array} \right)
\begin{pmatrix}
S_{1-N;p'\nu'}^{-s'}\\
S_{2-N;p'\nu'}^{-s'}\\
\vdots \\
S_{2N;p'\nu'}^{-s'}\\
S_{1-N;p'\nu'}^{+s'}\\
S_{2-N;p'\nu'}^{+s'}\\
\vdots\\
S_{2N;p'\nu'}^{+s'}
\end{pmatrix}
=s' \begin{pmatrix}
{\underline{{e}}}_{p'\nu'x}^{s'(\mathrm{m})}\\
{\underline{{e}}}_{p'\nu'y}^{s'(\mathrm{m})}\\
{\underline{{h}}}_{p'\nu'x}^{s'(\mathrm{m})}\\
{\underline{{h}}}_{p'\nu'y}^{s'(\mathrm{m})}\\
\end{pmatrix}  \label{eq:Smatrices_dynamic} \;,
\end{align}
\end{widetext}
for $p'=1,2$, $\nu'=-N,-N+1,\cdots,N$, and $s'=+(-)$. We note that $\mathrm{m}=1$ or 2 denotes the appropriate medium, which corresponds to $s'=+$ or $-$, respectively. 

In order to evaluate the scattering properties of multilayer structures, it	is convenient to express the waves on the left (right) side of a given interface about different points, at a distance $-\mathbf{d}_1$~($\mathbf{d}_2$) from a center at the interface, so that all backward and forward propagating (or evanescent) waves in medium m between two consecutive interfaces refer to the same (arbitrary) origin at $\mathbf{R}_\mathrm{m}$. 
This requires redefining the transmission and reflection coefficients of the interface by multiplying with the appropriate phase factors, as follows~\cite{comphy1,comphy2}
\begin{equation}\label{eq:qmatrices_dynamic}
\begin{split}
& Q_{p\nu;p'\nu'}^{\mathrm{I}}=  S_{p\nu;p'\nu'}^{++}\exp{\{i[\mathbf{q}_{p\nu}^{+(2)}\cdot\mathbf{d}_2+\mathbf{q}_{p'\nu'}^{+(1)}\cdot\mathbf{d}_1]\}}\\
& Q_{p\nu;p'\nu'}^{\mathrm{II}}= S_{p\nu;p'\nu'}^{+-}\exp{\{i[\mathbf{q}_{p\nu}^{+(2)}\cdot\mathbf{d}_2-\mathbf{q}_{p'\nu'}^{-(2)}\cdot\mathbf{d}_2]\}}\\
& Q_{p\nu;p'\nu'}^{\mathrm{III}}=S_{p\nu;p'\nu'}^{-+}\exp{\{-i[\mathbf{q}_{p\nu}^{-(1)}\cdot\mathbf{d}_1-\mathbf{q}_{p'\nu'}^{+(1)}\cdot\mathbf{d}_1]\}}\\
& Q_{p\nu;p'\nu'}^{\mathrm{IV}}= S_{p\nu;p'\nu'}^{--}\exp{\{-i[\mathbf{q}_{p\nu}^{-(1)}\cdot\mathbf{d}_1+\mathbf{q}_{p'\nu'}^{-(2)}\cdot\mathbf{d}_2]\}}\;.
\end{split}
\end{equation}

In the particular case of an interface between two static homogeneous media, Eq.~\eqref{eq:Smatrices_dynamic} is reduced to a set of $2N+1$ independent linear equations~\cite{christofi}
\begin{align}
\!\!\!\!&\left(\!\!
\begin{array}{ccccc} 
-{{{e}}}_{1nx}^{-(1)}&- {{{e}}}_{2nx}^{-(1)}&
 {{{e}}}_{1nx}^{+(2)}&  {{{e}}}_{2nx}^{+(2)}   \\
-{{{e}}}_{1ny}^{-(1)}&- {{{e}}}_{2ny}^{-(1)}&
 {{{e}}}_{1ny}^{+(2)}&  {{{e}}}_{2ny}^{+(2)}   \\
-{{{h}}}_{1nx}^{-(1)}& -{{{h}}}_{2nx}^{-(1)}& 
 {{{h}}}_{1nx}^{+(2)}&  {{{h}}}_{2nx}^{+(2)}   \\
-{{{h}}}_{1ny}^{-(1)}& -{{{h}}}_{2ny}^{-(1)}& 
 {{{h}}}_{1ny}^{+(2)}&  {{{h}}}_{2ny}^{+(2)}   \\
\end{array} \right)\!\!\!
\begin{pmatrix}
S_{1n;p'n}^{-s'}\\
S_{2n;p'n}^{-s'}\\
S_{1n;p'n}^{+s'}\\
S_{2n;p'n}^{+s'}
\end{pmatrix} \!\!=\!\!s' \!\!\begin{pmatrix}
{{{e}}}_{p'nx}^{s'(\mathrm{m})}\\
{{{e}}}_{p'ny}^{s'(\mathrm{m})}\\
{{{h}}}_{p'nx}^{s'(\mathrm{m})}\\
{{{h}}}_{p'ny}^{s'(\mathrm{m})}\\
\end{pmatrix} \;, \label{eq:Smatrices_iso_dynamic}
\end{align}
one for each value of $n=-N,-N+1,\cdots,N$. 
Therefore, the transmission and reflection matrices become diagonal in $n$, which reflects frequency conservation.

The transmission and reflection matrices of a pair of consecutive interfaces, $\mathrm{i}$ and $\mathrm{i}+1$, are obtained by properly combining those of the two interfaces so as to describe multiple scattering to any order. 
This leads to the following expressions after summing the infinite geometric series involved~\cite{comphy1,comphy2}

\begin{align}
		\nonumber \mathbf{Q}^{\mathrm{I}}&=
		\mathbf{Q}^{\mathrm{I}}_{(\mathrm{i}+1)} [\mathbf{I} \!-\!
		\mathbf{Q}^{\mathrm{II}}_{(\mathrm{i})}
		\mathbf{Q}^{\mathrm{III}}_{(\mathrm{i}+1)}]^{-1} \mathbf{Q}^{\mathrm{I}}_{(\mathrm{i})} \\
	\nonumber \mathbf{Q}^{\mathrm{II}} &=
	\mathbf{Q}^{\mathrm{II}}_{(\mathrm{i}+1)} +\ \mathbf{Q}^{\mathrm{I}}_{(\mathrm{i}+1)}
\mathbf{Q}^{\mathrm{II}}_{(\mathrm{i})} [ \mathbf{I} -
	\mathbf{Q}^{\mathrm{III}}_{(\mathrm{i}+1)}
	\mathbf{Q}^{\mathrm{II}}_{(\mathrm{i})}]^{-1} \mathbf{Q}^{\mathrm{IV}}_{(\mathrm{i}+1)} \\
	\nonumber \mathbf{Q}^{\mathrm{III}}&=
	\mathbf{Q}^{\mathrm{III}}_{(\mathrm{i})} \!+\! \mathbf{Q}^{\mathrm{IV}}_{(\mathrm{i})}
	\mathbf{Q}^{\mathrm{III}}_{(\mathrm{i}+1)} [\mathbf{I} \!-\!
	\mathbf{Q}^{\mathrm{II}}_{(\mathrm{i})}
	\mathbf{Q}^{\mathrm{III}}_{(\mathrm{i}+1)}]^{-1} \mathbf{Q}^{\mathrm{I}}_{(\mathrm{i})} \\
	\mathbf{Q}^{\mathrm{IV}}&=
		\mathbf{Q}^{\mathrm{IV}}_{(\mathrm{i})} [\mathbf{I} \!-\!
		\mathbf{Q}^{\mathrm{III}}_{(\mathrm{i}+1)}
		\mathbf{Q}^{\mathrm{II}}_{(\mathrm{i})}]^{-1} \mathbf{Q}^{\mathrm{IV}}_{(\mathrm{i}+1)} \label{eq:q_multiple} \;.
\end{align}

It is obvious that one can repeat the above process to obtain the transmission and reflection matrices $\mathbf{Q}$ of three consecutive interfaces, by combining those of the pair of the first interfaces with those of the third interface, and so on, by properly combining the ${Q}$ matrices of component units, one can obtain the ${Q}$ matrices of a slab which comprises any finite number of interfaces.

In general, we want to describe a multilayer slab, which comprises time-periodic media, sandwiched between two semi-infinite homogeneous and isotropic static media: (A) on its left and (B) on its right, characterized by scalar relative electric permittivities $\epsilon_\mathrm{A}$, $\epsilon_\mathrm{B}$ and magnetic permeabilities $\mu_\mathrm{A}$, $\mu_\mathrm{B}$, respectively.
The transmission and reflection coefficients of the structure are obtained using the pairing procedure described by Eq.~\eqref{eq:q_multiple}. 
For a wave incident on the slab from the left, of the form $\boldsymbol{\mathcal{E}}_{\mathrm{in}}(\mathbf{{r}},t)=\exp{\{i[\mathbf{q}_{n'}^{+\mathrm{(A)}}\cdot\mathbf{r}_\mathrm{A}+n'\Omega t)]\}}\mathbf{e}_{p'n'}^{+(\mathrm{A})}$, the transmitted and reflected fields are given by

\begin{align}
  \!\!\!\!\!\boldsymbol{\mathcal{E}}_\mathrm{tr}(\mathbf{r},t)\!\!&=\!\!\sum_{p,n}\! Q^\mathrm{I}_{pn;p'n'}\exp\big\{i[\mathbf{q}_{n}^{+\mathrm{(B)}}\!\cdot\!\mathbf{r}_\mathrm{B}+n\Omega t]\big\}{\mathbf{e}}_{pn}^{+\mathrm{(B)}}\!\!\!\!\\
  \!\!\!\!\!\boldsymbol{\mathcal{E}}_\mathrm{rf}(\mathbf{r},t)\!\!&=\!\!\sum_{p,n}\! Q^\mathrm{III}_{pn;p'n'}\exp\big\{i[\mathbf{q}_{n}^{-\mathrm{(A)}}\!\cdot\!\mathbf{r}_\mathrm{A}+n\Omega t]\big\}{\mathbf{e}}_{pn}^{-\mathrm{(A)}},\!\!
\end{align}
where $\mathbf{{r}}_\mathrm{m}=\mathbf{r}-\mathbf{R}_{\mathrm{m}}$, $\mathbf{q}_{n}^{\pm(\mathrm{m})}=\mathbf{q}_\parallel+q_{nz}^{\pm(\mathrm{m})}\widehat{\mathbf{z}}$, $q_{nz}^{\pm(\mathrm{m})}=\pm\sqrt{k_n^2\epsilon_{\mathrm{m}}\mu_{\mathrm{m}}-q_\parallel^2}$, with $\mathrm{m}=\mathrm{A,B}$ denoting the appropriate medium. 
We note that, in an isotropic medium such as A or B, the polarization eigenmodes are degenerate; this is why the polarization index $p$ (or $p'$) has been dropped from the corresponding wave vectors.
Having calculated the transmitted and reflected fields, we obtain the transmittance $\mathcal{T}$ and reflectance $\mathcal{R}$ of the slab. 
These are defined as the ratio of the flux of the transmitted or of the reflected field to the flux of the incident wave, respectively. 
Integrating the Poynting vector over the $x-y$ plane, on the each time appropriate side of the slab, and taking the average over a long time interval $\tau \rightarrow \infty$, we obtain 
\begin{align}
&  \mathcal{T}=\displaystyle\sum_{p,n}\mathcal{T}_{pn}=\sum_{p,n}   |{Q}_{pn;p'n'}^\mathrm{I}|^2 \dfrac{
	\mathrm{Re}[{{q}}_{nz}^{+(\mathrm{B})}]\mu_\mathrm{A}}{\mathrm{Re}[{{q}}_{n'z}^{+(\mathrm{A})}]\mu_\mathrm{B}} \label{eq:trans_dynamic}\\
& 
  \mathcal{R}=\displaystyle\sum_{p,n}\mathcal{R}_{pn}=\sum_{p,n}   |{Q}_{pn;p'n'}^\mathrm{III}|^2 \dfrac{
  	\mathrm{Re}[{{q}}_{nz}^{-(\mathrm{A})}]}{\mathrm{Re}[{{q}}_{n'z}^{+(\mathrm{A})}]}\;, \label{eq:refl_dynamic}
\end{align}
provided that $\omega\neq n\Omega/2$,~\cite{zurita} a condition that will be always satisfied in our case. 
Because of the time variation of the permittivity tensor, the EM energy is not conserved even in the absence of dissipative (thermal) losses. 
In this case, $\mathcal{A}=1-\mathcal{T}-\mathcal{R}>0 \,(<0)$ means energy tranfer from (to) the EM to (from) the spin-wave field. 

\subsection{Results} 
\begin{figure}
\centering
\includegraphics[width=\linewidth]{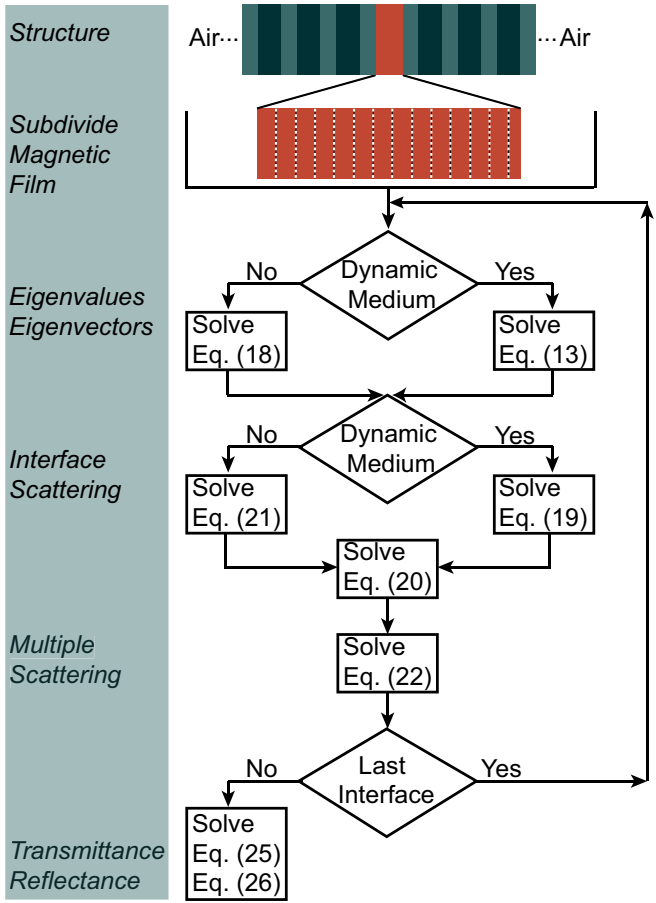}
\caption{ (Color online) Schematic description of the fully dynamic time-Floquet scattering-matrix method. }
\label{fig2}
\end{figure}
We shall now demonstrate the validity of the fully dynamic time-Floquet scattering-matrix method described above, and compare with the adiabatic approximation, on the structure of Fig.~\ref{fig1}(a).
We recall that this structure supports two neighboring defect modes in the lowest Bragg gap. 
As in Sec.~\ref{sc:structure}, we consider  p-polarized incident light, with frequency $\omega_0a/c=1.88375$ and $q_\parallel a=1.2$ that corresponds to the higher-frequency defect mode in Fig.~\ref{fig1}(c), and assume excitation of a second-order perpendicular standing spin wave of frequency $\Omega_2$ with a relative amplitude $\eta=0.1$.
The presence of the spin wave induces a spatial and time-periodic variation in the permittivity tensor [see Eq.~\eqref{eq:depsilon_decomp}].
To deal with spatial inhomogeneity we consider 50 elementary homogeneous sublayers in the magnetic film while the dynamic response is described assuming a truncation order $N=20$ in the Fourier series expansions involved. 
These parameters ensure excellent convergence of the numerical results.

Our method proceeds as schematically described in Fig.~\ref{fig2}. We first
calculate the eigenmodes of the EM field in each of the 53 (infinite)
homogeneous media involved, i.e., the 50 dynamic media that represent the
Bi:YIG film under the  action of the spin wave and the 3 static media
(air, $\mathrm{SiO}_{2}$, $\mathrm{TiO}_{2}$), solving the appropriate
eigenvalue-eigenvector equations, Eq.~\eqref{eq:system1} and Eq.~\eqref{eq:system_iso}, respectively.
Then, we evaluate the transmission and reflection matrices $\mathbf{Q}$ of
the consecutive interfaces by Eq.~\eqref{eq:Smatrices_dynamic}, or Eq.~\eqref{eq:Smatrices_iso_dynamic} in the simple case of
an interface between two static media, and Eqs.~\eqref{eq:qmatrices_dynamic}. At each step of the
iterative procedure, the $Q$ matrices of the entire slab built-up up to
the current interface are obtained according to Eqs.~\eqref{eq:q_multiple}. Therefore, at
the end of the procedure, the $Q$ matrices of the whole structure are
readily available and the corresponding transmittance and reflectance are
calculated from Eqs.~\eqref{eq:trans_dynamic} and~\eqref{eq:refl_dynamic}, respectively.

\begin{figure}
	\centering
	\includegraphics[width=1\linewidth]{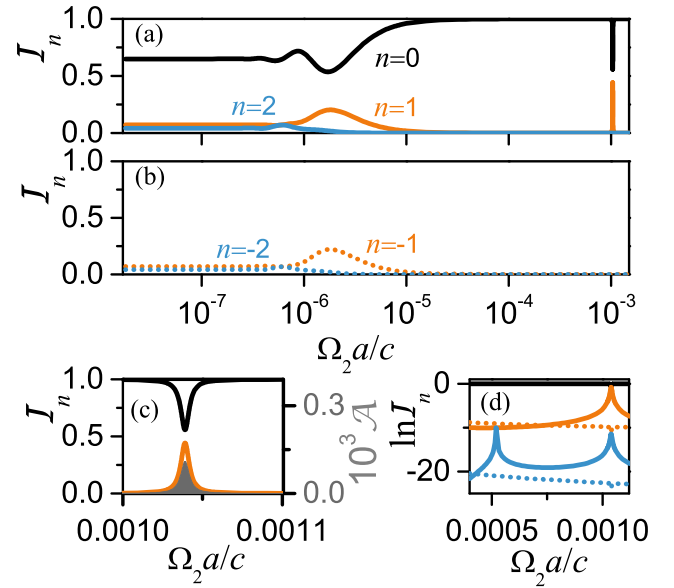}
	\caption{(Color online) The structure of Fig.~\ref{fig1} (a), under continuous excitation of a second-order perpendicular standing spin wave of frequency $\Omega_2$ with a relative amplitude $\eta=0.1$, is illuminated from the left by p-polarized light with $q_\parallel a=1.2$ at the resonance frequency $\omega_0 a/c=1.88375$. (a), (b) Variation of the elastic and inelastic total outgoing light intensities versus the spin-wave frequency $\Omega_2$ associated with emission (a) and absorption (b) of zero (black line), one (orange line), and two (blue line) magnons by a photon. (c), (d) An enlarged view in the region where the magnon(s) frequency matches the frequency difference between the two optical resonances shown in Fig.~\ref{fig1}(c) (triple-resonance condition). 
	The shaded area in (c) shows the corresponding optical absorption.}
	\label{fig3}
\end{figure}

Figure~\ref{fig3} displays the total (transmitted plus reflected) intensity of the elastic and inelastic outgoing light beams, associated with absorption and emission of zero, one, and two magnons by a photon, as a function of the spin-wave frequency. 
It can be seen that, at spin-wave frequencies smaller than the width of the optical resonance ($\Gamma a/c\sim10^{-6}$), we recover the results of the quasi-static adiabatic approximation. 
The calculated intensities of the elastic and inelastic outgoing light beams are in perfect agreement with those reported in Table~\ref{tab:intensities}, while the actual value of the spin-wave frequency is, indeed, immaterial.
In Figs.~\ref{fig3}(a) and (b) we see that, as $\Omega_2$ increases, the quasi-static adiabatic approximation breaks down and a fully dynamic approach, which properly takes into the account the initial and final photon states is required.
If $\Omega_2$ is of the order of the dynamical frequency shift of the optical resonance $(\Delta\omega a/c\sim10^{-6})$, the intensities  of the scattered light beams exhibit an oscillatory behavior. 
By further increasing $\Omega_2$, inelastic light scattering processes are suppressed, since the final states are within the gap, where the optical density of states is very low, and we are essentially left with only elastic scattering. 
At higher values of $\Omega_2$, the magnon frequency can match the frequency difference between the two optical resonances shown in Fig.~\ref{fig1}(c) (triple-resonance condition). 
In this case, one-magnon emission processes are favored, leading to enhanced intensities of the corresponding $n=1$ inelastically transmitted and reflected light beams, for both polarization-conserving and polarization-converting processes.
At the same time, elastic scattering is considerably reduced, by about 45\%, while the other inelastic processes are also resonantly affected, though to a much lower degree.
Overall, there is an excess number of magnons produced, which is manifested as a peak in the total optical absorption spectrum as depicted in Fig.~\ref{fig3}(c), thus clearly indicating a resonant energy transfer from light to spin waves. 
This effect, also, cannot be accounted for by the adiabatic approximation. 
It is worth-noting that the triple-resonance condition can be accomplished by many-magnon emission processes as well, however the effects are considerably weaker as shown in Fig.~\ref{fig3}(d).

So far, quantities are expressed in terms of the lattice parameter $a$.
Taking $a=300$~nm, the optical wave considered is tuned to a frequency $\omega_{0}/(2 \pi)=300$~THz, i.e., a wavelength $\lambda_{0}=1 \; \mu\mathrm{m}$. 
Correspondingly, the spin-wave frequency $\Omega_{2}/ (2 \pi)$ in Figs.~\ref{fig3}(a) and (b) ranges from about $10^{-2}$~GHz to $10^{2}$~GHz. 
It should be pointed out that the different features observed can be tuned in frequency by appropriately choosing the geometric parameters of the structure. 
For example, the triple-resonance condition can be shifted to lower frequencies, say to the few GHz region which is typical for perpendicular standing spin waves,~\cite{navabi2017} by increasing the lattice parameter $a$ and, accordingly, the optical wavelength. 
This will also require to use a lower Faraday coefficient,~\cite{zvezdin_book} leading to a smaller separation between the two optical resonances involved. We recall that the amplitude of the multiple-magnon processes increases with the coupling strength $\Delta \omega / \Gamma$. 
In this respect, it should be noted that the lifetime of a resonant defect mode ($\propto 1/ \Gamma$) can be made arbitrarily long by increasing the number of unit cells in the Bragg mirrors, provided that absorptive losses are negligible, while $\Delta \omega$, can be widened by raising the spin-wave amplitude.

\section{\label{conclusions}Conclusion}
In summary, we formulated a fully dynamic time-Floquet scattering-matrix method, with arbitrary-order accuracy, in order to describe  the interaction of visible and infrared light with spin waves in layered optomagnonic structures.
The method is robust, computationally efficient and, contrary to brute-force time-domain methods, is ideally suited for fast and accurate calculations of the optical response of periodically driven systems where two very different time scales are involved.
The applicability of the method is demonstrated on a specific model of a dual photonic-magnonic cavity, which operates in the strong-coupling regime. 
The cavity is formed by sandwiching a magnetic Bi:YIG film between two dielectric Bragg mirrors and is subject to continuous excitation of a perpendicular standing spin wave.
For spin-wave frequencies smaller than the width of the optical resonance, we recover the results of the adiabatic approach, namely enhanced modulation of light through multi-magnon absorption and emission mechanisms, beyond the linear-response approximation. 
For higher  spin-wave frequencies where the adiabatic approach breaks down, our fully dynamic calculations provide evidence for the occurrence of remarkable effects, such as enhanced inelastic light scattering in the nonlinear strong-coupling regime when a triple-resonance condition is fulfilled and resonant energy transfer between the photon and the magnon fields. 
 We note that similar effects should occur, also, for perpendicular standing spin waves in in-plane magnetized films as well as for surface Dammon-Eshbach and backward volume waves with an in-plane propagation wave vector.
In the latter cases, however, the spin wave induces a periodic variation of the electric 
permittivity tensor both in time and in the propagation direction. The implementation
of a space-time-Floquet scattering-matrix method for a fully dynamic description 
of such cases is currently in progress. 

\acknowledgments{Petros-Andreas Pantazopoulos is supported from
the General Secretariat for Research and Technology and the
Hellenic Foundation for Research and Innovation through a PhD
scholarship (No.~906).}

\appendix*
\section{}

In order to gain further insight into the accuracy and limits of validity of the quasi-static adiabatic approximation, let us consider a generic one-dimensional quantum mechanical model: A particle of mass $m$ in a periodically oscillating delta-function potential at the origin, in which case Schr\"{o}dinger's equation reads

\begin{equation} \label{eq:schroedinger}
 i \hbar \frac{\partial \Psi(x,t)}{\partial t} = - \frac{\hbar ^{2}}{2m}  \frac{\partial ^{2} \Psi(x,t)}{\partial x^{2}} +g \cos(\Omega t) \delta (x) \Psi(x,t)  \;.
\end{equation}
We note that this model was studied, though in a different perspective, also by Martinez and Reichl.~\cite{martinez}
We want to describe scattering of the particle, when it impinges with (kinetic) energy $E$ on the potential from $x<0$. 
It is straightforward to show from Eq.~(\ref{eq:schroedinger}) that the wave function of the incident particle, with amplitude equal to unity, is

\begin{equation}
   \Psi_{\mathrm{in}}(x,t)=\exp[i (k_{0}x-Et/\hbar)] \;,
\end{equation}
for $x<0$, where
\begin{equation}
   E= \frac{\hbar^{2} k_{0}^{2}}{2m} \;.
\end{equation}
We now assume that the variation of the potential is relatively slow, i.e., $\Omega \ll E/ \hbar \equiv \omega$, and undertake an adiabatic approach by solving the problem with the potential strength kept fixed and then, at the end of the calculation, allow it to change in successive snapshots within a period $T=2 \pi /\Omega$.
This implies transmitted and reflected waves of the form

\begin{equation}
   \Psi_{\mathrm{tr}}(x,t)=\frac{k_{0}}{k_{0}+img \cos( \Omega t)/\hbar^{2}}\exp[i (k_{0}x-\omega t)] \;,
\end{equation}
for $x>0$ and
\begin{equation}
   \Psi_{\mathrm{rf}}(x,t)=\frac{-img \cos( \Omega t)/\hbar^{2}}{k_{0}+img \cos( \Omega t)/\hbar^{2}}\exp[i (-k_{0}x-\omega t)] \;, 
\end{equation}
for $x<0$, which can be readily obtained by applying the well-known results for the static delta-function potential.~\cite{griffiths}

Expanding the periodically varying transmission and reflection coefficients into Fourier series we have

\begin{equation}
 \label{eq:trwave0}
  \!\!\!\Psi_{\mathrm{tr}}(x,t)=\!\!\!\!\!\!\!\!\sum_{n=0,\pm1,\pm2,\ldots} \!\!\!\!\!\!\!\!\!\!A^{0}_{\mathrm{tr}}(n) \exp \{i [k_{0}x-(\omega-n \Omega)t]\} \;,
\end{equation}
for $x>0$ and
\begin{equation}
 \!\!\!\Psi_{\mathrm{rf}}(x,t)=\!\!\!\!\!\!\!\!\sum_{n=0,\pm1,\pm2,\ldots} \!\!\!\!\!\!\!\!\!\!A^{0}_{\mathrm{rf}}(n) \exp \{i [-k_{0}x-(\omega-n \Omega)t]\} \;,    \label{eq:rfwave0}
\end{equation}
for $x<0$, where
\begin{equation}
   A^{0}_{\mathrm{tr}}(n)=\frac{1}{T} \int_{0}^{T} dt \frac{k_{0} }{k_{0}+img\cos (\Omega t)/\hbar^{2}} \exp(-i n \Omega t)
\end{equation}
and $A^{0}_{\mathrm{rf}}(n) = A^{0}_{\mathrm{tr}}(n)-\delta_{n0}$. 
The Fourier coefficients $A^{0}_{\mathrm{tr}}(n)$ can be evaluated analytically by contour integration, which yields

\begin{equation}
   \label{eq:adiabatic}
   A^{0}_{\mathrm{tr}}(n)=\frac{\alpha_{0}}{\sqrt{\alpha_{0}^{2}+1}} i^{|n|}(\alpha_{0} - \sqrt{\alpha_{0}^{2}+1})^{|n|}\;,
\end{equation}
where $\alpha_{0} \equiv k_{0} \hbar^{2}/(mg)$, and the actual value of $\Omega$ is of course immaterial.

Equation~(\ref{eq:trwave0}) and~(\ref{eq:rfwave0}) tell us that the transmitted and reflected wave functions consist of an infinite number, $n=0, \pm1,\pm2, \ldots$, of plane wave beams of the same wave number $k_{0}$, frequencies $\omega - n \Omega$, and amplitudes $A^{0}_{\mathrm{tr}}(n)$ and $A^{0}_{\mathrm{rf}}(n)$, respectively. 
The corresponding transmittance, $\mathcal{T}$, and reflectance, $\mathcal{R}$, are defined as the time-averaged [$\lim_{\tau \longrightarrow \infty} \frac{1}{\tau} \int_{0}^{\tau} dt (\cdots)$] probability current [$\mathbf{J}= \frac{\hbar}{m} \mathrm{Im}(\Psi^{*} \nabla \Psi)$] associated with the transmitted and reflected waves in the direction of flow, $\widehat{\mathbf{x}}$ and $-\widehat{\mathbf{x}}$, respectively, relative to that of the incident wave. 
Using Eqs.(\ref{eq:rfwave0}) we obtain

\begin{equation}
   \label{eq:trans0} 
   \mathcal{T}= \sum_{n=0,\pm1,\pm2,\ldots} |A^{0}_{\mathrm{tr}}(n)|^{2} \equiv \sum_{n=0,\pm1,\pm2,\ldots} \mathcal{T}_{n}
\end{equation}   
\begin{equation}
   \label{eq:refl0} 
   \mathcal{R}= \sum_{n=0,\pm1,\pm2,\ldots} |A^{0}_{\mathrm{rf}}(n)|^{2} \equiv \sum_{n=0,\pm1,\pm2,\ldots} \mathcal{R}_{n} \;.
\end{equation}
We note that, considering the Fourier series expansions $\alpha_{0}/(\alpha_{0}+ i \cos \phi) = \sum_{n=0,\pm1,\pm2,\ldots} A^{0}_{\mathrm{tr}}(n) \exp(i n \phi)$ and $-i \cos \phi/(\alpha_{0}+ i \cos \phi) = \sum_{n=0,\pm1,\pm2,\ldots} A^{0}_{\mathrm{rf}}(n) \exp(i n \phi)$, Parseval's theorem implies $\mathcal{T} +\mathcal {R} =1$, as expected for a nondissipative adiabatically changing system.

On the other hand, a fully dynamic solution to the scattering problem described above can be obtained by a Fourier expansion of the incident, transmitted, and reflected wave functions, prior to applying the appropriate boundary conditions. 
We have

\begin{align}
\!\!\Psi_{\mathrm{in}} (x,t) &=\!\!\!\!\!\!\!\! \!\!\!\sum_{n=0,\pm1,\pm2,\ldots} \!\!\!\!\!\!\!\! A_{\mathrm{in}}(n) \exp \{i [k_{n}x-(\omega-n \Omega)t]\} \;, \nonumber \\
\!\!\Psi_{\mathrm{tr}} (x,t) &=\!\!\!\!\!\!\!\! \!\!\!\! \sum_{n=0,\pm1,\pm2,\ldots} \!\!\!\!\!\!\!\! A_{\mathrm{tr}}(n) \exp \{i [k_{n}x-(\omega-n \Omega)t]\} \;,  \nonumber \\
\!\!\Psi_{\mathrm{rf}} (x,t) &= \!\!\!\!\!\!\!\!\!\!\!\! \sum_{n=0,\pm1,\pm2,\ldots} \!\!\!\!\!\!\!\!A_{\mathrm{rf}}(n) \exp \{i [-k_{n}x-(\omega-n \Omega)t]\} \;, 
\end{align}

where
\begin{equation}
   \hbar (\omega-n \Omega) = \frac{\hbar^{2} k_{n}^{2}}{2m} \;.
\end{equation}
Continuity of the total wave function $\Psi(x,t)$ ($\Psi=\Psi_{\mathrm{in}}+ \Psi_{\mathrm{rf}}$ for $x<0$ and $\Psi=\Psi_{\mathrm{tr}}$ for $x>0$) requires

\begin{equation}
   \label{eq:cont1} 
   A_{\mathrm{in}}(n) + A_{\mathrm{rf}}(n) - A_{\mathrm{tr}}(n) = 0 \;, 
\end{equation}
for $ n=0,\pm 1,\pm 2, \ldots$. Moreover, integrating Eq.~(\ref{eq:schroedinger}), we obtain a discontinuity in the first derivative of the total wave function

\begin{equation}
   \frac{\partial \Psi(x,t)}{\partial x}\bigg|_{x\rightarrow 0^{+}} - \frac{\partial \Psi(x,t)}{\partial x}\bigg|_{x\rightarrow 0^{-}} = \frac{2m}{ \hbar^{2}}g \cos(\Omega t) \Psi(0,t) \;,
\end{equation}
which, with the help of Eq.~(\ref{eq:cont1}), leads to

\begin{equation}
   \label{eq:cont2} 
   A_{\mathrm{tr}}(n) + \frac{i}{2 \alpha_{n}} A_{\mathrm{tr}}(n-1) + \frac{i}{2 \alpha_{n}} A_{\mathrm{tr}}(n+1) = A_{\mathrm{in}}(n) \;,
\end{equation}
for $n=0,\pm 1,\pm 2, \ldots$, where $\alpha_{n} \equiv k_{n} \hbar^{2}/(mg)$.
\begin{figure} [t]
   \centering
   \includegraphics[width=1\linewidth]{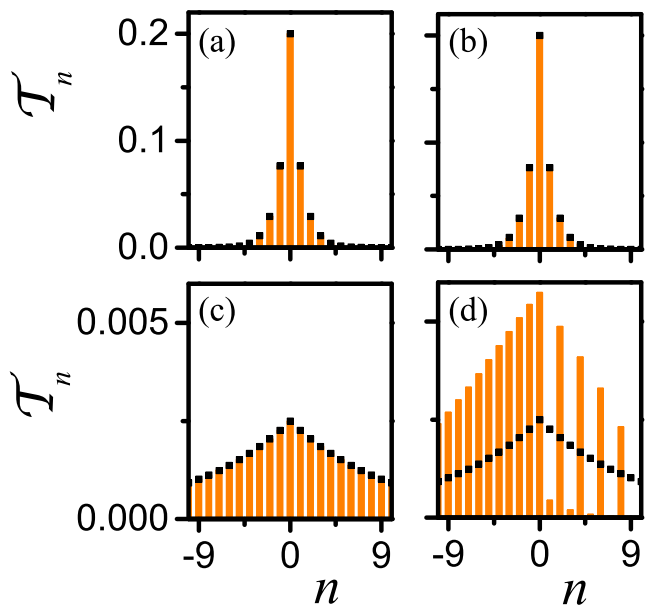}
   \caption{(Color online) Partial transmittance associated with the different plane wave beams, $\mathcal{T}_{n}$, calculated by the adiabatic approximation (points) and by the fully dynamic  approach (orange bars), for different values of $\alpha_{0}$ and $\Omega/ \omega$. (a): $\alpha_{0}=0.5$, $\Omega/\omega=10^{-3}$, (b): $\alpha_{0}=0.5$, $\Omega/\omega=10^{-1}$, (c): $\alpha_{0}=0.05$, $\Omega/\omega=10^{-3}$, and (d): $\alpha_{0}=0.05$, $\Omega/\omega=10^{-1}$. } 
   \label{fig4}
\end{figure}
Equation~(\ref{eq:cont2}) provides the fully dynamic solution for the transmitted wave function while the reflected wave function can be readily obtained from Eq.~(\ref{eq:cont1}). 
The corresponding transmittance and reflectance are given by
\begin{align}
   &\mathcal{T}= \sum_{n=0,\pm1,\pm2,\ldots} \frac {|A_{\mathrm{tr}}(n)|^{2} \; \mathrm{Re} (k_{n})}{k_0} \equiv \sum_{n=0,\pm1,\pm2,\ldots} \mathcal{T}_{n} \nonumber\\
  \label{eq:refl} 
  &\mathcal{R}= \sum_{n=0,\pm1,\pm2,\ldots} \frac {|A_{\mathrm{rf}}(n)|^{2} \; \mathrm{Re} (k_{n})}{k_0} \equiv \sum_{n=0,\pm1,\pm2,\ldots} \mathcal{R}_{n} \;.
\end{align}
We note that here, contrary to the quasi-static adiabatic description, for $n> \omega / \Omega$ we obtain evanescent plane wave beams that do not contribute to the transmittance and the reflectance because $\mathrm{Re} (k_{n})=0$.

It would be now interesting to compare the results of the adiabatic approximation with those of the fully dynamic description in the limit $\Omega \ll \omega$. 
Taking $\alpha_{n} = \alpha_{0}$, Eq.~(\ref{eq:cont2}), for $A_{\mathrm{in}}(n)=\delta_{n0}$ as appropriate to the case under consideration, yields directly $A_{\mathrm{tr}}(n)$ equal to the middle ($n=0$) column of the inverse of a tridiagonal Toeplitz matrix with the elements of its main diagonal equal to unity and the elements of its first diagonals below and above the main diagonal equal to $i/(2 \alpha_{0})$. 
Using the explicit expressions for the inverse of a general tridiagonal Toeplitz matrix of dimensions $(2N+1) \times(2N+1)$,~\cite{fonseca} we obtain in our case here

\begin{equation}
   \label{eq:dynamic} 
   A_{\mathrm{tr}}(n)=(-1)^{|n|}(-2i \alpha_{0})
\frac{U_{N-|n|}(-i \alpha_{0}) U_{N}(-i \alpha_{0})} {U_{2N+1}(-i\alpha_{0})} \;,
\end{equation}
 for $n=0, \pm 1, \pm 2, \ldots, \pm N $, where $U_{p}(z)$ are the Chebyshev polynomials of the second kind~\cite{abramowitz_book} given by

\begin{equation}\label{eq:chebyshev_general}
   U_{p}(z)= \frac{r_{+}^{p+1}-r_{-}^{p+1}}{r_{+}-r_{-}}
\end{equation}
with $r_{\pm}=z \pm \sqrt{z^{2}-1}$. 
Interestingly, restricting to $|n|\ll N$, in the limit of matrix dimensions tending to infinity, Eq.~(\ref{eq:dynamic}) leads strictly to Eq.~(\ref{eq:adiabatic}): $A_{\mathrm{tr}}(n)=A^{0}_{\mathrm{tr}}(n)$.

For $\alpha_{n} \neq \alpha_{0}$, the matrix to be inverted is a tridiagonal Jacobi (non-Toeplitz) matrix and an elegant, concise formula for its inverse is also available.~\cite{fonseca}
Though in this general case no simple analytic solution can be obtained, considering $\Omega /\omega \leq 10^{-3}$, numerical calculations show that the absolute relative difference between the results of the adiabatic approximation and the fully dynamic approach does not exceed $10^{-3}$. 
The calculated partial transmittance associated with the different plane wave beams, $\mathcal{T}_{n}$, considering sufficiently large tridiagonal matrices to ensure convergence, is depicted in Fig.~\ref{fig4} for some characteristic cases. 
It can be seen that strong deviations from the results of the adiabatic approximation appear only for simultaneously large-amplitude ($\alpha_{0} = 0.05$) and fast ($\Omega / \omega =0.1$) variations. 
It is worth-noting that, in quantum mechanics, the transmittance and reflectance usually refer to probability currents, as in the example considered here, and not to energy currents as commonly described in electromagnetism by the Poynting vector.~\cite{jackson} 
Probability current conservation ($\mathcal{T} + \mathcal {R} = 1$) is here always satisfied (to machine accuracy $\sim 10^{-16}$) in both adiabatic and fully dynamic calculations, as expected for the Hermitian Hamiltonian system under study. 
In addition to that, the adiabatic approximation should also preserve energy current conservation, i.e., since particles in the $n$-th beam have an energy $\hbar (\omega - n \Omega)$, the quantity $\sum_{n=0,\pm1,\pm2,\ldots} n \Omega (\mathcal{T}_{n} + \mathcal{R}_{n})$ must vanish identically, which is indeed ensured by the symmetry properties $\mathcal{T}_{-n} = \mathcal{T}_{n}$ and $\mathcal{R}_{-n} = \mathcal{R}_{n}$ implied by Eq.~(\ref{eq:adiabatic}). 
Of course, these symmetry properties are not satisfied in the fully dynamic description of the problem, as can be clearly seen in the bottom right diagram of Fig.~\ref{fig4}, while small deviations from the perfectly symmetric partial transmittance also exist in the other diagrams but they are not discernible in the scale of the figure. 
This is translated to the absence of energy conservation, as expected in the actual  time-varying system under consideration. 
It becomes, therefore, clear from the example elaborated above that the quasi-static adiabatic approximation provides a reasonably good description of wave scattering in systems with a relatively slow (but not necessarily weak) periodic variation with time, at least to the extent where the scattering spectral features do not exhibit strong changes over a frequency range of the order of $\Omega$ such as in inelastic scattering between resonant modes.

\end{document}